\documentclass[letter, 10pt, conference]{./ieeeconf}
\makeatletter
\def\endthebibliography{%
  \def\@noitemerr{\@latex@warning{Empty `thebibliography' environment}}%
  \endlist
}
\makeatother

\IEEEoverridecommandlockouts 
\IEEEtriggeratref{15}

\usepackage{times} 
\usepackage{float} 
\usepackage{enumerate} 
\usepackage{amsmath} 
\usepackage{mathrsfs}  
\usepackage{bm}
\usepackage{amssymb}  
\usepackage{xcolor}
\usepackage{optidef}
\usepackage{glossaries}
\usepackage{booktabs}
\usepackage[font=footnotesize]{caption}
\usepackage{subcaption}
\usepackage{algpseudocode}
\usepackage{tikz}

\usepackage[compress]{cite}

\makeatletter
\let\NAT@parse\undefined
\makeatother
\usepackage[pagebackref=false,breaklinks=true,colorlinks=true,bookmarks=false,allcolors=black,citecolor=black]{hyperref}

\usepackage{graphicx}
\graphicspath{{figures/}}
\usepackage[acronym,style=super,nogroupskip,nonumberlist,toc=false]{glossaries-extra}
\setabbreviationstyle[acronym]{long-short}
\newacronym[description={model predictive control(ler)}]{mpc}{MPC}{model predictive control}
\newacronym{com}{CoM}{center of mass}
\newacronym{rhc}{RHC}{receding horizon control}
\newacronym{pid}{PID}{proportional-integral-derivative}
\newacronym{lq}{LQ}{linear quadratic}
\newacronym{lqr}{LQR}{linear quadratic regulator}
\newacronym{dare}{DARE}{discrete algebraic ricatti equation}
\newacronym{dae}{DAE}{differential-algebraic equation}
\newacronym{ode}{ODE}{ordinary differential equation}
\newacronym{dp}{DP}{dynamic programming}
\newacronym{lp}{LP}{linear program}
\newacronym{nlp}{NLP}{nonlinear program}
\newacronym{qp}{QP}{quadratic program}
\newacronym{zoh}{ZOH}{zero-order hold}
\newacronym{rk4}{RK4}{fourth order Runge--Kutta}
\newacronym{lti}{LTI}{linear time-invariant}
\newacronym{jit}{JIT}{just-in-time}
\newacronym{dof}{DOF}{degrees of freedom}
\newacronym{api}{API}{application programming interface}
\newacronym{ros}{ROS}{Robot Operating System}
\usepackage{cleveref}
\Crefname{figure}{Fig.}{Figs.}
\Crefname{equation}{Eq.}{Eqs.}
\usepackage{physics}
\usepackage[mode=text]{siunitx}

\usepackage[utf8]{inputenc}
\usepackage[T1]{fontenc}
\usepackage[USenglish]{babel}

\newcommand{\eg}{\textit{e.g.}}

\usepackage{amsthm}

\theoremstyle{plain}

\usepackage{listings}
\makeatletter
\def\lst@makecaption{%
  \def\@captype{table}%
  \@makecaption
}
\makeatother

\newcommand\copyrighttext{%
  \footnotesize © 2024 IEEE.  Personal use of this material is permitted.  Permission from IEEE must be obtained for all other uses, in any current or future media, including reprinting/republishing this material for advertising or promotional purposes, creating new collective works, for resale or redistribution to servers or lists, or reuse of any copyrighted component of this work in other works.}
\newcommand\copyrightnotice{%
\begin{tikzpicture}[overlay, remember picture]
\node[anchor=south,yshift=10pt] at (current page.south) {\fbox{\parbox{\dimexpr\textwidth-\fboxsep-\fboxrule\relax}{\copyrighttext}}};
\end{tikzpicture}%
}

\title{\LARGE \bf
MPC-CBF with Adaptive Safety Margins for Safety-critical Teleoperation over Imperfect Network Connections
}

\author{Riccardo Periotto\textsuperscript{$\dagger$},
Mina Ferizbegovic\textsuperscript{$\ddagger$},
Fernando S. Barbosa\textsuperscript{$\ddagger$}, and
Roberto C. Sundin\textsuperscript{$\ddagger$}
\thanks{
\textsuperscript{$\dagger$} Riccardo Periotto is with Agile Robots AG, Munich, Germany.
\textsuperscript{$\ddagger$}Mina Ferizbegovic, Fernando S. Barbosa, and Roberto C. Sundin are with Ericsson Research, Stockholm, Sweden.
        E-Mail: {\tt\small \{mina.ferizbegovic, fernando.dos.santos.barbosa, roberto.castro.sundin\}@ericsson.com}}
}

\makeatletter
\def\footnoterule{\relax%
  \kern-5pt
  \hbox to \columnwidth{\vrule width 0.5\columnwidth height 0.4pt\hfill}
  \kern4.6pt}
\makeatother

\begin{document}

\maketitle
\thispagestyle{empty}
\pagestyle{empty}

\begin{abstract}
The paper focuses on the design of a control strategy for safety-critical remote teleoperation. The main goal is to make the controlled system track the desired velocity specified by an operator while avoiding obstacles despite communication delays. 
Control Barrier Functions (CBFs) are used to define the safety constraints that the system has to respect to avoid obstacles, 
while Model Predictive Control (MPC) provides the framework for adjusting the desired input, taking the constraints into account. The resulting input is sent to the remote system, where appropriate low-level velocity controllers translate it into system-specific commands.
The main novelty of the paper is a method to make the CBFs robust against the uncertainties caused by the network delays affecting the system's state and do so in a less conservative manner.
The results show how the proposed method successfully solves the safety-critical teleoperation problem, making the controlled systems avoid obstacles with different types of network delay.
The controller has also been tested in simulation and on a real manipulator, demonstrating its general applicability when reliable low-level velocity controllers are available.
\end{abstract}
\section{Introduction}\copyrightnotice
\emph{Teleoperation} is the technical term indicating the remote control of systems over long distances.
Over the last decades, it has been successfully applied in various scenarios, including nuclear power plants, submarine and space missions, medical operations, and industrial procedures \cite{vertut_teleoperation_1985}.
The number of applications in which it acts as a baseline is continuously growing. 
In particular, its integration with robotics has led to the development of a specialized field named \emph{telerobotics}, which aims to merge the objectives of both disciplines \cite{siciliano_springer_2008}.

In telerobotics, two fundamental elements to consider to ensure correct and reliable behavior of the system are \emph{control} and \emph{safety}, where the former is the one in charge of making the system achieve certain tasks, while the latter of respecting specific constraints \cite{vertut_teleoperation_1985, zeng_challenges_2022}.
In recent years, the relevance of safety has grown considerably, particularly due to the increasing adoption of \emph{autonomous systems}. By definition, autonomous systems are machines capable of performing tasks without human intervention.
In this type of application, safety is no longer solely an operator's responsibility but becomes a requirement that the system has to fulfill by itself. 
Typical examples of safety requirements are those needed to comply with regulations, guarantee the protection of people working nearby, and avoid damaging the environment. 

\begin{figure}[t]
  \begin{center}
    \includegraphics[width=0.45\textwidth]{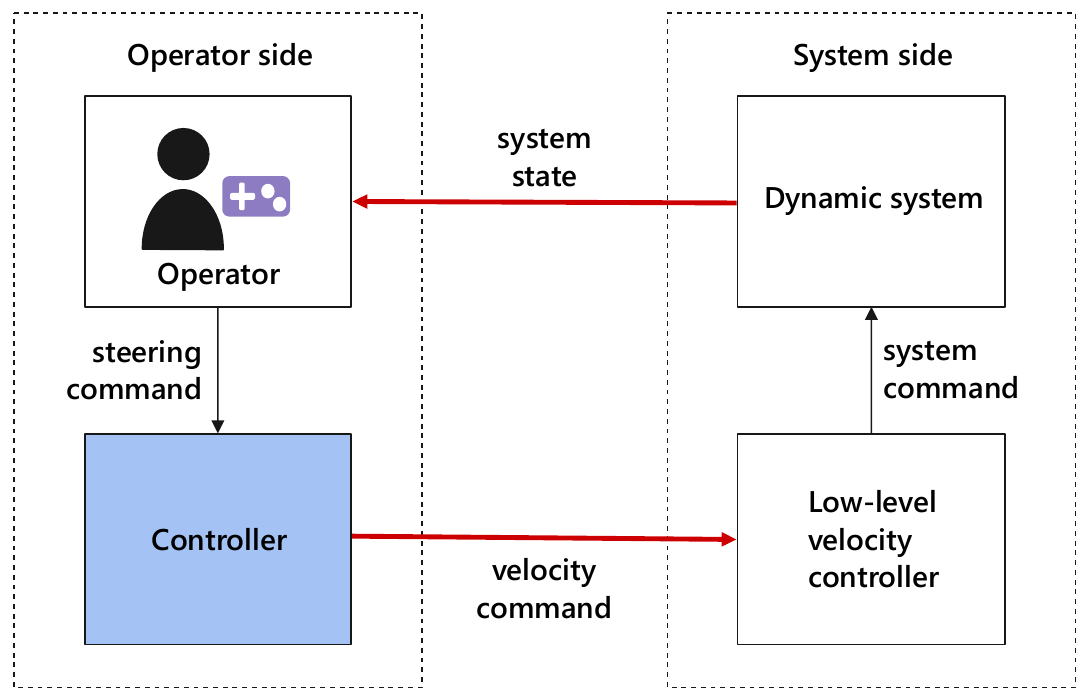}
  \end{center}
  \caption[High-level problem architecture]{
    \textbf{High-level problem architecture.} 
    The two larger dashed boxes represent the two sides of the network. The controller executes on the same side as the operator controlling the system, while the system and the low-level controller are on the opposite side. 
    The thicker red arrows represent the information shared between the two sides through the remote connection, while the thinner ones represent the information exchanged between elements operating in the same location. 
    The highlighted block symbolizes the main focus of the paper, which is an optimization-based controller in charge of modifying the desired velocity specified by the operator to enhance the safety guarantees of the system.
  }
  \label{fig:problem_architecture}
\end{figure}

The problem addressed in this work is the safety-critical teleoperation of dynamic systems.
The high-level architecture showing the entities involved in this problem and their interactions is presented in 
\Cref{fig:problem_architecture}, where an operator
uses a steering device (e.g., a gamepad) to control the system.
The signal of the steering device is encoded as a desired velocity 
and sent to a controller module in charge of refining it considering the safety constraints and sending it through a network connection.
This controller is the part of the architecture on which this paper primarily focuses.
Once the velocity command reaches the system's side, low-level controllers translate it to specific system's inputs.
The operator guides the system based on the received representation of it. This can be, for example, a video stream 
captured with one or several cameras or a real-time rendering in dedicated software.

This paper concentrates on the challenges related to the corruption of information due to network delay over remote connections and proposes a method to account for the resulting system uncertainties.
More precisely, we explore the effectiveness of approaches combining Model Predictive Control (MPC) and Control Barrier Functions (CBFs) for obstacle avoidance when the teleoperation is affected by communication delays.
Our contributions are twofold: i) we provide a real-time capable controller based on MPC and CBF that allows for online obstacle avoidance and; ii) propose a dynamic safety margin that adapts to network performance such that the system becomes resilient towards network disturbances. The adaptable safety margin makes the method less conservative than approaches based on upper bounds and worst-case scenarios.
The paper is structured as follows: \Cref{sec:RelatedWork} contextualizes the content of the paper within the existing literature, while \Cref{sec:Preliminaries} presents the theory underlying our method. \Cref{sec:Methods} explains the method and \Cref{sec:Results} illustrates the results, together with how we collected them. \Cref{sec:Conclusions} concludes the paper.

\section{Related Work}
\label{sec:RelatedWork}
The problem of controlling robots while avoiding obstacles is historically known. Offline trajectory planning is a possible method to address this problem, but it is only effective when both the environment and the path the system has to follow are known \cite{siciliano_springer_2008, hu_nmpc-mp_2021}. 
In many real-world applications, the environment where the system operates and the desired trajectory are unknown in advance, making online solutions essential.
Among the relevant and traditionally known online approaches, those based on haptic feedback and Artificial Potential Fields (APFs) play a significant role. Nonetheless, they do not provide the safety guarantees given by techniques such as CBFs \cite{singletary_comparative_2020, ferraguti_safety_2022}.

First introduced in \cite{wieland_constructive_2007} and later refined and popularized in \cite{ames_control_2019, ames_control_2014, ames_control_2017}, CBFs have now become a popular tool in the control literature. They are an extension of Barrier Functions (BFs) to systems with control inputs and can transform constraints defined on the system state to constraints on its inputs \cite{li_optimizable_2022}.
Multiple sources attribute their success to their ability to ensure safety, obstacle avoidance, and invariance properties, as well as their natural relationship with Lyapunov functions \cite{ames_control_2014, ames_control_2017, wu_safety-critical_2015}.
Indeed, just as Control Lyapunov Functions (CLFs) provide a mathematical framework for stability analysis, CBFs provide a complementary framework for safety.
In many works, they act as a safety regulator by minimally modifying the input calculated by a stabilizing or tracking controller. 

The main limitations of a standard application of CBFs such as the one presented in \cite{ames_control_2014}, \cite{ames_control_2017}, and \cite{xu_safe_2018}, are that: they are defined in continuous time, they modify the input by only considering the information available without any predictive approach, and they do not take into account the uncertainties of the system's state.
Recent approaches attempt to overcome these shortcomings by adopting mechanisms to: (i) perform in discrete time, (ii) consider the future behaviors of the system, and (iii) increase the tolerance against disturbances.

Of these mechanisms, the first has been adopted widely in the field, with several works also demonstrating its properties mathematically \cite{agrawal_discrete_2017, singletary_control_2020}.

The second mechanism can be implemented by considering a model of the system's dynamics within the optimization problem used to compute the input. In this regard, solutions based on MPC and their variants are increasingly being proposed, rather than more standard approaches based on Quadratic Programming (QP) \cite{rubagotti_semi-autonomous_2019, hu_nmpc-mp_2021, sankaranarayanan_paced-5g_2023}.
However, a significant limitation that the proposed work share is the adoption of Euclidean norms to define the obstacle avoidance constraints, for example by requiring the distance between the robot and the obstacles to be larger than a given value \cite{zeng_safety-critical_2020}.
The drawback of this approach is that it restricts the system's movement only when it is relatively close to the obstacles, technically, when the reachable set along the horizon intersects with the obstacles \cite{zeng_safety-critical_2020}.
Consequently, to enable the agent to take proactive actions to avoid obstacles, a longer prediction horizon is required in these cases, increasing the complexity of the optimization problem \cite{zeng_safety-critical_2020, zeng_challenges_2022, li_optimizable_2022}.
The definition of constraints through CBFs partially mitigates the problem, but undermines the invariance and feasibility properties provided by other MPC implementations.

The third mechanism relates more closely to the definition of safety constraints through CBFs and how to make them robust in the presence of uncertainties or external disturbances, which may degrade or compromise the safety guarantees \cite{alan_disturbance_2023}.
To address this challenge, researchers have proposed extensions of CBFs defined considering the uncertainties and disturbances affecting the system using online parameter adaptation \cite{taylor_adaptive_2020}, data-driven approaches \cite{lopez_robust_2021}, and disturbance-observer-based techniques \cite{alan_disturbance_2023}. 
The research in this field of robustifying CBFs is still active and new approaches are continuously being proposed depending on the scenario considered \cite{alan_parameterized_2023}.

The method proposed in this paper originates from the combination of these three mechanisms, with CBF constraints robustified with a time-varying safety margin proportional to the network disturbance, modifying the commanded control input in a discrete and predictive fashion.

\section{Preliminaries}
\label{sec:Preliminaries}
\subsection{Robust CBFs}
\label{sec:robust-cbfs}
Let us consider the nominal dynamics of a (possibly) nonlinear, control-affine system 
\begin{equation}
\label{eq:nominal_system}
    \dot{\mathbf{x}}(t) = f_c(\mathbf{x}(t))+g_c(\mathbf{x}(t))\mathbf{u}(t)\text{,}
\end{equation}
where $\mathbf{x}(t)\in\mathbb{R}^n$ and $\mathbf{u}(t)\in\mathcal{U}_\text{adm}\subset\mathbb{R}^m$ denote the state and the input of the system, and $\mathcal{U}_\text{adm}$ the admissible set of inputs. 
Functions $f_c:\mathbb{R}^n\rightarrow \mathbb{R}^n$ and $g_c:\mathbb{R}^n\rightarrow \mathbb{R}^{n\times m}$ are assumed to be locally Lipschitz continuous functions, while the input corresponds to a feedback controller of the form $\mathbf{u}(t)=k(\mathbf{x}(t))$, such that the resulting dynamical system in \Cref{eq:nominal_system} also is locally Lipschitz.

A CBF is then commonly defined as any continuously differentiable function 
$h:\mathbb{R}^n\to\mathbb{R}$ for which there exists an extended class $\mathcal{K}_\infty$ function $\alpha$ such that for all $\mathbf{x}\in\mathbb{R}^n$:
\begin{equation}
    \label{eq:cbf_condition}
    \sup_{\mathbf{u}\in\mathcal{U}_\textup{adm}}\qty[
    L_{f_c}h(\mathbf{x})+L_{g_c}h(\mathbf{x})\mathbf{u}
    ]\geq - \alpha(h(\mathbf{x})),
\end{equation}
where $L_{f_c}h(\mathbf{x})=\nabla h(\mathbf{x})f_c(\mathbf{x})$ and $L_{g_c}h(\mathbf{x})=\nabla h(\mathbf{x})g_c(\mathbf{x})$ are Lie derivatives of $h$ along $f_c$ and $g_c$, see \cite{ames_control_2019, lopez_robust_2021, perko_differential_2013, alan_parameterized_2023}.

Let us now consider an extension of the nominal dynamics
\begin{equation}
    \label{eq:uncertain_system}
    \dot{\mathbf{x}}(t)=f_c(\mathbf{x}(t))+g_c(\mathbf{x}(t))\mathbf{u}(t)+\tilde{f}(\mathbf{x}(t))+\tilde{g}(\mathbf{x}(t))\mathbf{u}(t),
\end{equation}
where the new functions $\tilde{f}:\mathbb{R}^n\rightarrow\mathbb{R}^n$ and $\tilde{g}:\mathbb{R}^n\rightarrow\mathbb{R}^{n\times m}$ are locally Lipschitz continuous and represent the differences between the real and nominal system models \cite{alan_parameterized_2023, jankovic_robust_2018}.
A possible approach to ensure safety despite model mismatch is to introduce a robustifying term $\sigma(\mathbf{x},\mathbf{u})$ in the safety constraint. 
Taking inspiration from \cite{jankovic_robust_2018}, a continuously differentiable function $h$ is defined in \cite{alan_parameterized_2023} as a Robust CBF (RCBF) for \Cref{eq:uncertain_system} on the 0-superlevel set of $h$, if $0$ is a regular value of $h$\footnote{A point $q$ is a \emph{regular value} of $h$ if $h(x)=q$ implies $\nabla h(x)\neq 0$.} and there exist functions $\sigma:\mathbb{R}_{\geq 0}\times\mathbb{R}~\rightarrow~\mathbb{R}$ and $\alpha\in\mathcal{K}_{\infty}$ such that
\begin{equation}
    \label{eq:rcbf_condition}
    \sup_{\mathbf{u}\in\mathcal{U}_\textup{adm}}[\dot{h}_n(\mathbf{x},\mathbf{u})-\sigma(\mathbf{x},\mathbf{u})] \geq -\alpha(h(\mathbf{x})),
\end{equation}
with $h_n$ being the CBF for the nominal dynamics as in (\ref{eq:cbf_condition}).

The idea behind introducing $\sigma$ is to make the controller take a more conservative action against the undesired effects on safety caused by uncertainties. In \cite{alan_parameterized_2023}, the authors illustrate an example of robust safety in a case with only bounded additive uncertainty (\textit{i.e.}, $||\tilde{f}(\cdot)||\leq \bar{w}$ and $\tilde{g}(\cdot)=0$). As we will present, our method follows the same setting, assuming that the uncertainties affecting the system result in a state shift representable as an additive term.

\subsection{CBF discretization}
\label{sec:cbf-discretization}
When defining CBFs, a popular choice is to choose $\alpha(\cdot)$ as a linear function with constant coefficient $\beta\in\mathbb{R}_{\geq 0}$, \textit{i.e.} $\alpha(h(\mathbf{x}))~=~\beta h(\mathbf{x})$. Assuming this choice and following the reasoning in \cite{li_optimizable_2022}, we can approximate $\dot{h}$ of \Cref{eq:cbf_condition} with
\begin{equation}
    \label{eq:h_derivative}
    \dot{h}(\mathbf{x}, \mathbf{u})\approx\frac{\Delta h}{\Delta t}=\frac{h_{t+\Delta t}-h_t}{\Delta t}\geq -\beta h_t,
\end{equation}
where $h_{t+\Delta t}$ and $h_t$ represent the value of $h(\mathbf{x})$ in the next and in the current time step, while $\Delta t$ is the control time interval.
Following \cite{agrawal_discrete_2017}, \Cref{eq:h_derivative} can be reformulated and extended to the discrete-time domain as 
\begin{equation}
    \label{eq:cbf_condition_discrete}
    \Delta h(\mathbf{x}_k, \mathbf{u}_k)\geq \gamma h(\mathbf{x}_k)
\end{equation}
where $\gamma\in(0,1]$ is a new variable incorporating $\Delta t$ and $\beta$, while $\Delta h(\mathbf{x}_k, \mathbf{u}_k)=h(\mathbf{x}_{k+1})-h(\mathbf{x}_k)$. 
With this, \Cref{eq:cbf_condition_discrete} is also equivalent to
\begin{equation}
\label{eq:dcbf_condition}
    h(\mathbf{x}_{k+1}) \geq (1-\gamma)h(\mathbf{x}_k),
\end{equation}
which is the condition playing the role of \Cref{eq:cbf_condition} in the case of discrete CBFs (DCBFs).

\section{Methods}
\label{sec:Methods}
Our method aims at unifying MPC and CBF for the purpose of increasing safety over a networked control architecture.
For this purpose, there are a few things to consider to form our MPC problem: (i) the continuous-time nature of standard CBFs,
(ii) the lack of precise real-time information of the system state due to network delays, and (iii) the lack of knowledge of
the operator's behavior throughout the entire MPC prediction horizon.

For (i), we follow the discretization steps of \Cref{sec:cbf-discretization} to arrive at
\begin{equation}
\label{eq:odcbf_condition}
    h(\mathbf{x}_{k+1}) \geq \omega(1-\gamma)h(\mathbf{x}_{k}),
\end{equation}
where we have added an optimizable slack variable $\omega\in\mathbb{R}_{\geq 0}$ to the CBF in \Cref{eq:dcbf_condition}.
The slack variable $\omega$ allows relaxing the decay rate imposed by the CBF, thereby enhancing feasibility. 
However, in order to minimize the deviation from the nominal decay rate $1-\gamma$, a regularization term, such as $\Phi(\omega):=P_{\omega}(\omega-1)^2$ for a $P_{\omega}\in\mathbb{R}_{> 0}$, should be added to the cost function \cite{zeng_enhancing_2021} (or see \Cref{problem:mpc-cost_function}). 

For (ii), we let our MPC rely on a predicted value of the state obtained by integrating the system dynamic model over an estimation of the network delay, such as seen in, \eg, \cite{sankaranarayanan_paced-5g_2023}.
Predicting the behavior of the operator, (iii), would depend upon the operator we are considering. If the operator is a high-level controller or trajectory planner, it is possible to know a priori the desired trajectory.
In the case of a human operator, this is, however, not the case. For this work, we opted for a method that would be agnostic to the type of operator, and thus followed the approach proposed in \cite{chipalkatty_less_2013}, according to which \emph{the simplest prediction methods outperform the more complex ones}. 
In practical terms, we let the MPC adopt the last velocity input received as the desired velocity throughout the entire horizon.

To deal with the uncertainties due to delays, we propose the addition of a safety margin $\sigma_k$, analogous to \Cref{eq:rcbf_condition}, to the discrete time
CBF in \Cref{eq:odcbf_condition}, obtaining
\begin{equation}
    \label{eq:odrcbf_condition}
    h(\mathbf{x}_{k+1}) -\sigma_k \geq \omega_k(1-\gamma)h(\mathbf{x}_{k}).
\end{equation}
Inspired by \cite{alan_parameterized_2023,zeng_challenges_2022,li_optimizable_2022},
we call the function satisfying this condition Optimal Decay RCBF (ODRCBF). 
However, unlike $\sigma(\mathbf{x}, \mathbf{u})$ in \Cref{eq:rcbf_condition}, we further propose
that $\sigma_k$ should also be a function of a set of network quality measures, such as
round-trip time (RTT), packet loss frequency, etc.

After the above derivations, we arrive at an MPC optimization problem of the form\footnote{ $\norm{v}_M$ represents the magnitude of vector $v$ weighted with the positive matrix $M$; mathematically $\norm{v}_M=\sqrt{v^\top M v}$.}:
\begin{subequations}
    \label{problem:mpc}
    \begin{align}
     J_k^*
    =&\min_{\mathbf{u}_{k:k+N-1}, \mathbf{\Omega_{1:n_{\text{obs}}}}}p(\mathbf{u}_{k+N}, \hat{\mathbf{u}}_k^\text{des})   \nonumber \\ 
     +&\sum_{i=0}^{N-1}\qty[q(\mathbf{u}_{k+i}, 
     \hat{\mathbf{u}}_k^\text{des}) + 
     \sum_{j=1}^{n_{\text{obs}}}\Phi(\omega^{(j)}_{k+i})] 
     \label{problem:mpc-cost_function} \\
    \text{s.t.: } & \mathbf{x}_{k+i+1}=f_{\text{kin}}(\mathbf{x}_{k+i}, \mathbf{u}_{k+i})
    \nonumber \\
        & \mathbf{x}_{k+i} \in \mathcal{X} 
        \nonumber \\
        & \mathbf{u}_{k+i} \in \mathcal{U}_\text{adm} 
        \nonumber \\
        & \mathbf{x}_{k} = \hat{\mathbf{x}}_k  
        \nonumber \\
        &  h^{(j)}(\mathbf{x}_{k+i+1}) -\sigma_{k} \geq \omega^{(j)}_{k+i}(1-\gamma)h^{(j)}(\mathbf{x}_{k+i}) \label{problem:mpc-safety_constraints} \\
        & \mathbf{\Omega}_j=\omega^{(j)}_{k:k+N-1} \nonumber\\
        & \omega^{(j)}_{k+i} \geq 0 \nonumber
    \end{align}
\end{subequations}
where $p(\mathbf{u}_{k+N}, \hat{\mathbf{u}}_k^\text{des}) = \norm{\mathbf{u}_{k+N}-\hat{\mathbf{u}}_k^\text{des}}^2_P$,
$q(\mathbf{u}_{k+i}, \hat{\mathbf{u}}_k^\text{des})=\norm{\mathbf{u}_{k+i}-\hat{\mathbf{u}}_k^\text{des}}^2_Q + \norm{\mathbf{u}_{k+i}}^2_R + \norm{\Delta\mathbf{u}_{k+i+1}}^2_{S}$, and $\Phi$ is the previously introduced
regularization function. 
$N$ is the horizon length and  $i\in\{0,\ldots,N-1\}$, while $j\in \{1,\ldots, n_{\text{obs}}\}$, with $n_{\text{obs}}$ being the number of obstacles.
$\mathbf{\Omega}_j$ denotes the slack variables associated with obstacle $j$.
$P$, $Q$, $R$, and $S$ are positive definite weight matrices. 
$f_{\text{kin}}:\mathbb{R}^n\times\mathbb{R}^m\to\mathbb{R}^n$ denotes the system kinematics;
$\hat{\mathbf{u}}_k^\text{des}$ is the predicted desired input, while $\hat{\mathbf{x}}_k$ the predicted state (both at timestep $k$); 
$\mathcal{X}$ is the state domain, which corresponds to the 3D space in case there are no constraints, otherwise it is the working space the 
controlled system is subject to. Note that the state domain does not need to take into account the obstacles as those are considered directly
by the ODRCBFs. As usual for MPC, only the first entry of the optimal solution $\mathbf{u}^*_{k} = \arg\min J_k^*$ is applied to the system.

A major difference from the standard formulation is that the safety constraints in \Cref{problem:mpc-safety_constraints} are defined through ODRCBFs, rather than through Euclidean distances or other types of CBFs as done in previous works.
Note that a constraint like the one referenced has to be integrated for each obstacle and timestep related to the problem definition.

Another difference comes from our assumption on the existence of a low-level controller (see \Cref{fig:problem_architecture}) so that the objective is to track a desired velocity, rather than a path.

Now that we have described the main parts of the formulation, we can dig into the details and explain how we define the $h$ function and the safety margin $\sigma$ in our specific case.

In this paper, we approximate both the robot and the obstacles as spheres.
In addition, we assume we want to have a minimum distance margin $d_{\min}$ between the robot and the obstacle. 
The rationale behind this margin is that, even if the system and the obstacle do not touch each other, conditions become unsafe when the distance between the two is below a certain threshold. The margin can also be applied to cases when the true state of the system is subject to uncertainties.
Given $\mathbf{x}$ and $\mathbf{p}^{(j)}$ as the centers of the robot and of obstacle $j$, $r_{\text{rob}}$ and $r^{(j)}_{\text{obs}}$ their spherical radii, and $d_{\min}$ the minimum distance margin, we define the function $h^{(j)}$ as 
\begin{equation}
    h^{(j)}(\mathbf{x})=\norm{\mathbf{x}-\mathbf{p}^{(j)}_{\text{obs}}}^2 - \qty(r_{\text{rob}} + r^{(j)}_{\text{obs}} + d_{\min})^2.
\end{equation}

Concerning $\sigma$, we already stated that its objective is to make the controller filter the velocity 
command provided and compensate for the uncertainties affecting the system's information. 
As mentioned, we only assume additive uncertainty caused by delays, which translates to a shift between the 
system's position used by the controller at a given time and the real one. 
This shift is proportional to both the delays and the system's velocity.
There are multiple ways to combine these variables to design a safety margin. The approach we propose adopts the following equation
\begin{equation}
  \label{eq:sigma} \sigma_k=\norm{\mathbf{v}_k}\cdot\bar{T}^\text{RTT}_k+v_{\max}\cdot k_\sigma \cdot \bar{\sigma}^{\text{RTT}}_k,
\end{equation}
where $\norm{\mathbf{v}_k}$ and $v_{\max}$ are the system's velocity and maximum velocity magnitudes, while $\bar{T}^\text{RTT}_k$ and $\bar{\sigma}^{\text{RTT}}_k$ the estimated values of the round trip time (RTT) and its standard deviation.
The idea is to make the safety margin directly proportional to the delay affecting the information and its oscillation, with the possibility of adjusting it by changing the system's velocity or the factor $k_{\sigma}\in\mathbb{R}_{\geq 0}$. 
Note that, in a scenario such as the one we are considering, there are multiple sources of delay: the delay of transmitting the input, the computation time for solving the optimization problem, and the delay of receiving the system state information. Despite this, their effect on the shift described above is similar and we can thus consider them all together by
using the RTT.

Given the above, we can see our method based on the $\sigma_k$ defined in \Cref{eq:sigma} as another data-driven approach to calculate a time-varying safety margin for RCBFs similar to those reported in \cite[Table 1]{alan_parameterized_2023}, but in the case of a system disturbed by delays.

\section{Results}
\label{sec:Results}
\begin{figure}
    \centering
    \vspace{0.2cm}
    \includegraphics[width=\columnwidth]{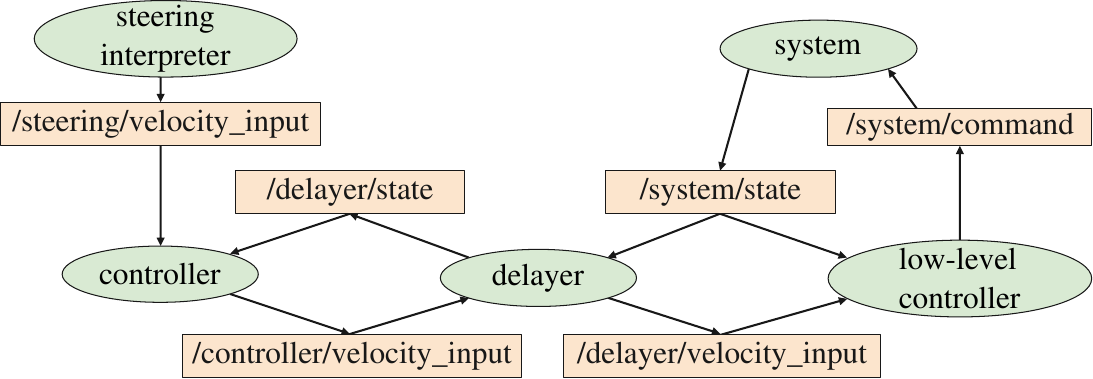}
    \caption[Node architecture]{\textbf{Node architecture.} Graph showing the connections between the different entities involved in the project. Each entity corresponds to a ROS~2 node and is represented with an ellipse, while the topics are in rectangles. The graph has been redrawn from that obtained using the rqt\_graph tool.}
  \label{fig:rqt_graph}
\end{figure}
\begin{figure}
  \begin{center}
    \includegraphics[width=0.45\textwidth]{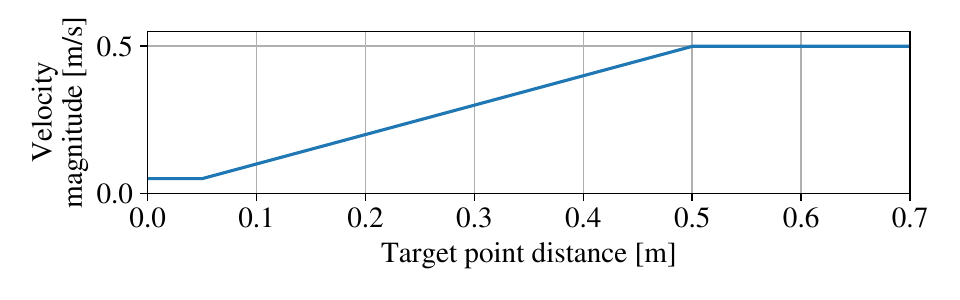}
  \end{center}
  \caption[Desired velocity magnitude law]{
        \textbf{Desired velocity magnitude law.} Relationship between the desired velocity magnitude specified by the tester and the system's distance from the target position.
        When the distance is smaller than \SI{0.15}{m}, its magnitude is at the lower limit \SI{0.05}{m/s}. On the other hand, when the distance is greater than \SI{0.50}{m}, the magnitude is at its upper limit \SI{0.50}{m/s}.
     }
  \label{fig:desired_velocity_shape}
\end{figure}
\begin{figure*}
     \centering
     \begin{subfigure}[b]{0.553\columnwidth}
         \centering
         \includegraphics[width=\textwidth]{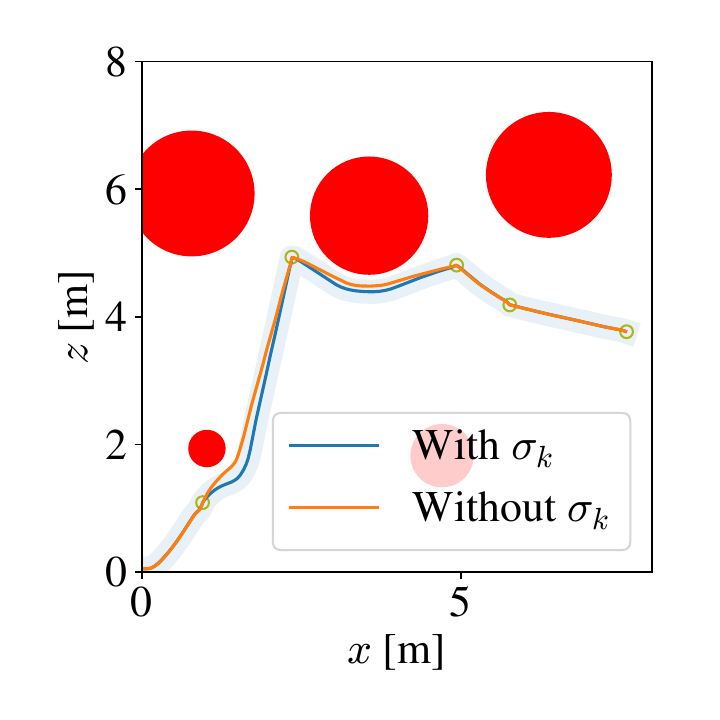}
         \caption{}
         \label{fig:task5_gaussian_xz}
     \end{subfigure}
     \begin{subfigure}[b]{0.7133\columnwidth}
         \centering
         \includegraphics[width=\textwidth]{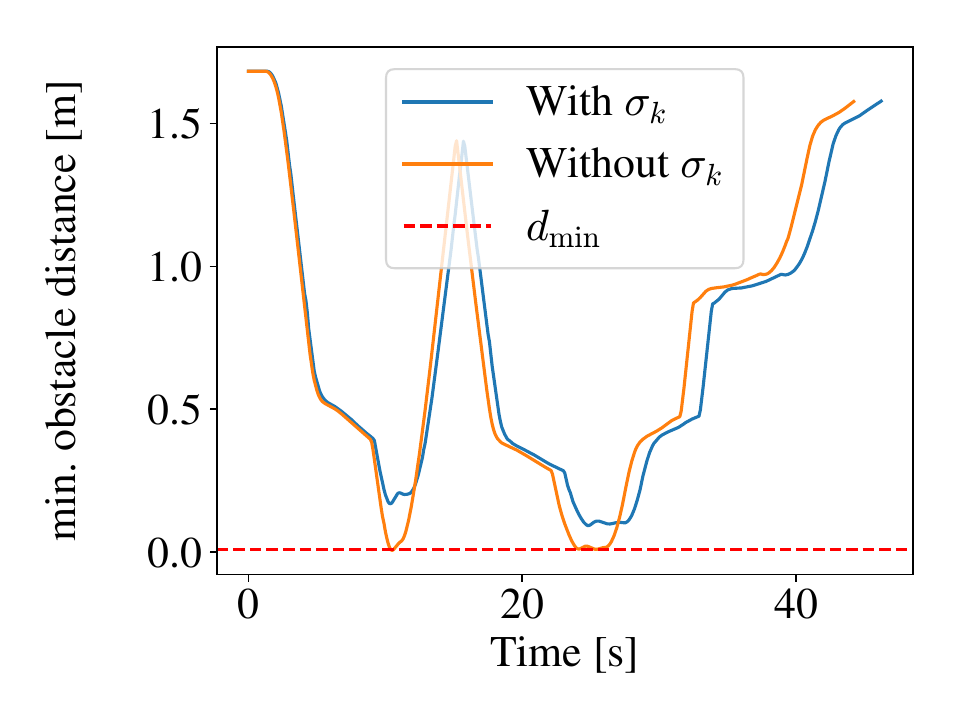}
         \caption{}
         \label{fig:task5_gaussian_distance}
     \end{subfigure}
     \begin{subfigure}[b]{0.7133\columnwidth}
         \centering
         \includegraphics[width=\textwidth]{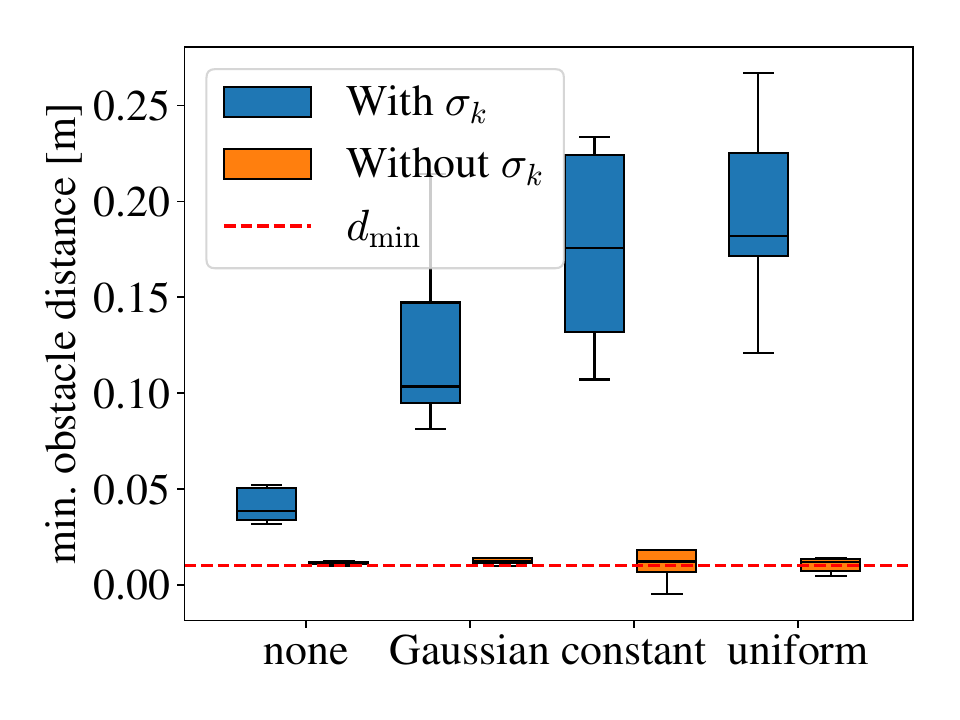}
         \caption{}
         \label{fig:box_plot_min_dist}
     \end{subfigure}
     \caption[Simulation results]{
        \textbf{Simulation results.} (a) $xz$-plane projections of the paths followed by the system for a task in simulation with obstacles
        in red and the target points as yellow circles.
        The communication is affected by a Gaussian delay with \SI{50}{ms} mean and \SI{20}{ms} standard deviation. (b) A minimum distance to obstacle for both methods and the same task. (c) Statistical results for all tasks showing minimum obstacle distance. The whiskers denote the most extreme values.
     }
     \label{fig:simulation_results}
\end{figure*}
\begin{figure}
     \centering
     \begin{subfigure}[b]{0.65\columnwidth}
         \centering
         \includegraphics[width=\textwidth]{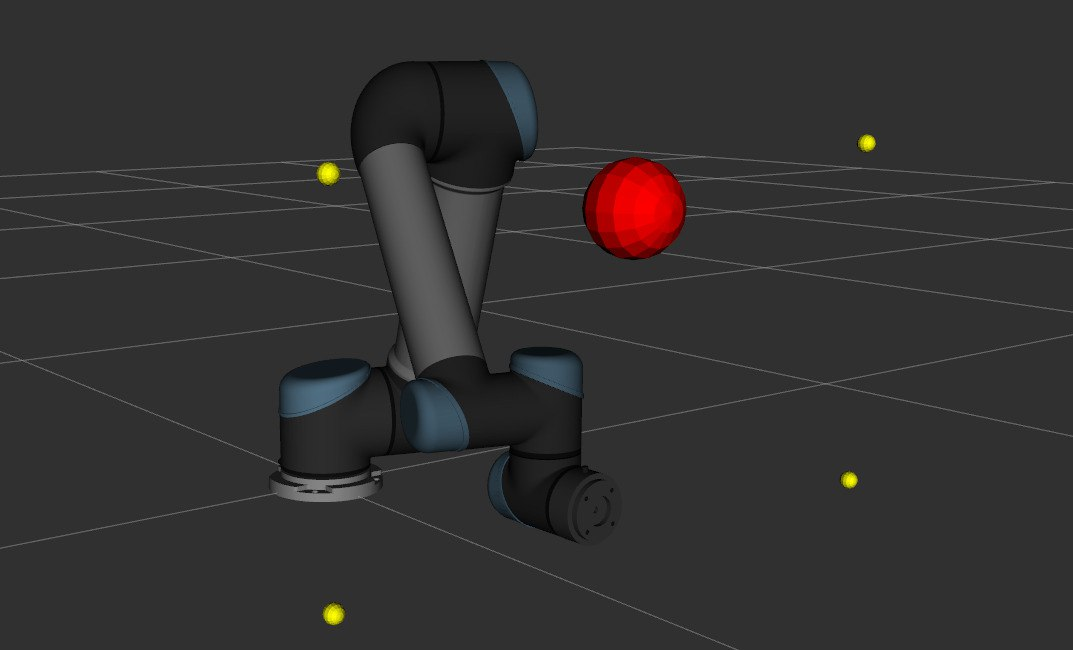}
         \caption{}
         \label{fig:hardware_rendering}
     \end{subfigure}
     \begin{subfigure}[b]{0.2465\columnwidth}
         \centering
         \includegraphics[width=\textwidth]{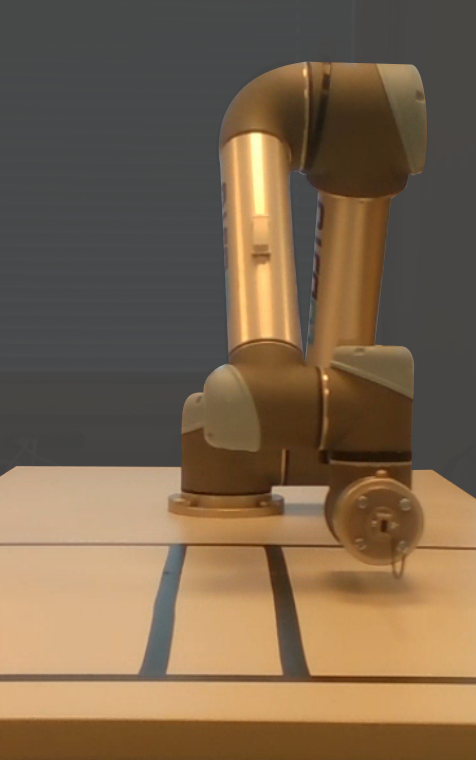}
         \caption{}
         \label{fig:ur5-real}
     \end{subfigure}
    \caption[Task example.]{
        \textbf{Hardware setup.} 
        (a) 3D rendering of the UR5 manipulator and the target task in rviz. The red sphere represents the obstacle the end effector has to avoid, while the yellow ones target points the end effector has to reach.
        (b) The real UR5 manipulator used in the experiment.
     }
     \label{}
\end{figure}

In this section, we aim to validate the controller presented in \Cref{sec:Methods} in simulation and a hardware setup. In simulation, we evaluate the performance difference in a few metrics when the controller utilizes the proposed dynamic safety margin $\sigma_k$ against the case when it does not ($\sigma_k =  0$). In hardware, we validate that the controller with safety margin $\sigma_k$ can run in real time and that the safety constraints are not violated when the system operates under realistic network delay.

For both the simulated and real case, we use ROS~2 Galactic \cite{ros2} as the underlying communication infrastructure with a setup as illustrated in \Cref{fig:rqt_graph}, displaying the 
rqt\_graph\footnote{\url{http://wiki.ros.org/rqt_graph}} with the connections between the different ROS nodes and topics.
The nodes at its core are: (i) the steering interpreter translating the steering commands sent by the operator into 3D Cartesian velocity input, (ii) the low-level velocity controller, (iii) a delayer node capable of simulating network delays in packet transmission, (iv) a system node which receives low-level control commands and sends its state, and (v) our controller.
The delayer node essentially acts as an intermediary between the two sides of the communication and delays the transmission of messages by holding them for a given time. 
The delay values can be either generated from a probability distribution or taken from a dataset of delay samples.

To ensure an unbiased comparison of various controller instances, we conducted the experiments using what we refer to as the \emph{tester} module. This module sends the velocity commands to make the system 
complete the task, both in the simulation and hardware scenarios.
Its role is to mimic a human operator, directing the system toward its next target position, but without considering the obstacles. The relationship between the velocity command sent by the tester module and the system's distance from the target position is depicted in \Cref{fig:desired_velocity_shape}.
Note that, it is not crucial that the behavior just described represents a real operator's intention; the crucial part for our considerations is to maintain a consistent method for imposing identical velocities under the same conditions, allowing for a fair comparison of the different formulations.
\begin{figure}
     \centering
         \includegraphics[width=0.9\columnwidth]{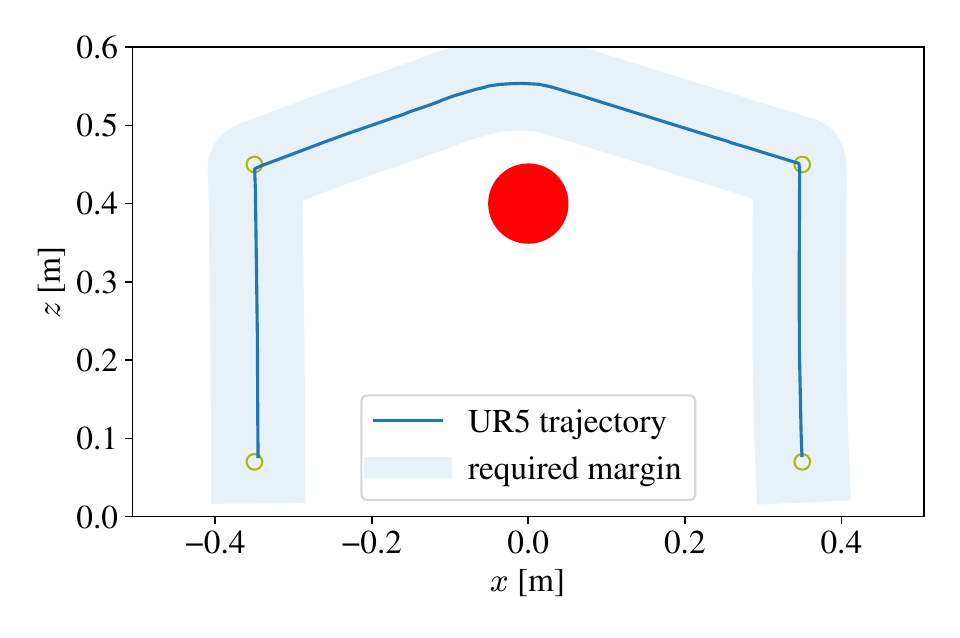}
     \caption[]{
     \textbf{Hardware results.} 
     Results from the hardware experiment using the UR5 robot with a recorded set of network delays of an
     average RTT of \SI{11.61}{ms}. The obstacle is marked in red, and the target points are yellow circles.
     The minimum required margin $r_{\text{obs} + d_{\min}}$ has been highlighted in light blue along
     the UR5 trajectory.
     }
     \label{fig:ur5_details}
\end{figure}

The tasks assigned to the two systems in the two scenarios were different, but based on the same idea
of reaching a set of target points in the working space while avoiding obstacles. The location of 
target points and obstacles were generated depending on the experiment.

The proposed controller \Cref{problem:mpc} was 
implemented as a CasADi \cite{casadi} model with Ipopt \cite{ipopt} as a solver, and
initially tested in a simulated environment using Gazebo\footnote{\url{https://gazebosim.org/home/}} and later on a Universal Robots UR5\footnote{\url{https://universal-robots.com/products/}} robotic manipulator operating in the laboratory. 
We used the following values for matrices in \Cref{problem:mpc}:
\begin{equation}
    Q=\mathit{I}_3\cdot10\text{,  }
    R=\mathit{I}_3\text{,  }
    P=\mathit{I}_3\cdot10\text{,  }
    S=\mathit{I}_3\cdot10\text{,  }
    P_\omega=100\text{,} \nonumber
\end{equation}
and let $d_{\min} = \SI{0.01}{m}$ and $\gamma = 0.5$.
The system kinematics were set using the Euler forward method such that
$\hat{\mathbf{x}}_{k+i+1} = \hat{\mathbf{x}}_{k+i} + \dd{t}\mathbf{u}_{k+i}$, where $\dd t$ was set to \SI{0.1}{s}.
The MPC was running at a frequency of \SI{10}{Hz} in both cases and with a horizon $N = 10$. 
The values $\bar{T}^\text{RTT}$ and $\bar{\sigma}^{\text{RTT}}_k$ were estimated using the moving average approach,
with a window length of 30 samples. 
We let $k_\sigma=1$ for both cases, while  $\norm{v_{\text{max}}}$ is set to \SI{0.87}{m/s} and \SI{0.52}{m/s} for the simulation
and hardware experiment, respectively.
While no constraints were set on the state, we set input constraints
such that for any element $u$ in $\mathbf{u}_{k:k+N-1}$, $\abs{u}\leq u_{b}$, where
$u_b$ was set to 
\SI{0.5}{m/s} in simulation and
\SI{0.3}{m/s} in the hardware experiment.
\subsection{Simulation results}
\label{subsec:Simulation}
In the simulation, the system controlled is a free-flyer box capable of moving in 3D space.
The box is \SI{0.25}{m} wide, \SI{0.25}{m} long, and \SI{0.1}{m} high and has a mass of \SI{1}{kg}.
We thus let $r_{\text{rob}}$ be defined as the radius of the circumscribed sphere of this cuboid.
The simulation framework used was Gazebo 11, and velocity control of the system was achieved by coupling the 
Gazebo Force plugin with a PID controller, and the system position and velocity was given through the 
Gazebo P3D plugin.

In our evaluations, 10 different tasks were constructed by placing out a random number of obstacles and target points in the $xz$-plane in the domain $[0.45\text{ m}, 7.6\text{ m}]^2\times[0,0]$. The position of the obstacles and target points, and the radius of the obstacles were random, but to ensure feasibility, we resampled if an obstacle or target point was too close to another obstacle.
For the delay, we considered four cases: (i) no delay, (ii) delay from a Gaussian distribution with \SI{50}{ms} mean and \SI{20}{ms} standard deviation, (iii) constant delay of \SI{200}{ms}, and (iv) delay from a uniform distribution over the interval 50-\SI{200}{ms}.

\Cref{tab:qualitative_comparison} shows the success rate obtained by the proposed method in simulation, when guiding the system to complete the 10 tasks with and without the safety margin $\sigma_k$, and for the different delay distributions. The 90 \% success rate for the case of no delay and without $\sigma_k$, is likely to be attributed to the fact that, even if there are no network delays affecting the communication,
the information is always subject to the delay caused by the time needed to solve the optimization problem, which can be mitigated through the safety margin.

\begin{table}
  \vspace{1.0em}
  \begin{center}
    \caption[Aggregated simulation results.]{\textbf{Aggregated simulation results.} Success rate of the proposed method with and without the safety margin $\sigma_k$ aggregated by delay type. The results refer to the execution of 10 different tasks in the simulation.}
    \label{tab:qualitative_comparison}
    \begin{tabular}{l|cccc}
      \textbf{Delay type} & 
      \textbf{None} & 
      \textbf{Gaussian} &
      \textbf{Constant} & 
      \textbf{Uniform} \\
      Params [ms] & 
      - & 
      $\mu=50\text{, }\sigma=20$ &
      $200$ &
      $[50-20]$ \\
      \hline
      With $\sigma_k$ & 100\% & 100\% & 100\% & 100\%\\
      Without $\sigma_k$ & 90\% & 90\% & 60\% & 70\%\\
    \end{tabular}
  \end{center}
\end{table}

To better demonstrate the effect of the safety margin, one of the tasks subject to Gaussian delay is shown in 
\Cref{fig:task5_gaussian_xz,fig:task5_gaussian_distance}. \Cref{fig:task5_gaussian_xz} shows a $xz$-plane 
projection of the results, while \Cref{fig:task5_gaussian_distance} shows the minimum distance to an obstacle 
over time. Clearly, the execution becomes unsafe at times when the safety margin $\sigma_k$ is not used, while
the controller with $\sigma_k$ maintains the distance well above $d_{\min}$ at all times, even though the overall
trajectory is minimally modified. The safety margin is thus not leading to an overly conservative behavior, but
rather, it makes the controller intervene only when the system is close to reaching an unsafe region.
\Cref{fig:box_plot_min_dist} shows the statistics for all ten tasks and
indicates that the increased ability to maintain a safe distance when using $\sigma_k$ is a general trend.
Further, we notice that this also applies for the case of no delay, which we again attribute to the increased
safety added by taking the computation time into account.
\subsection{Hardware results}
\label{subsec:UR5_res}
The hardware system we used to test the controller is a Universal Robots UR5 manipulator operating in the Kista laboratory as seen in \Cref{fig:ur5-real}. Unlike the previous case, the velocity commands obtained by the optimizer and sent through the network to the low level controller are not directly applicable to this type of system, but have to be translated into joint motor commands. We utilized Moveit Servo to handle this translation, allowing seamless execution of the desired motion on the UR5 manipulator end effector while also providing end effector position estimates.
The task tested on this system is similar to that of the previous case,
although the distances between the points are generally shorter because of the limitation of the working space reachable by the robot.
For this experiment we let $r_{\text{rob}}=\SI{5}{cm}$, always positioned 
at the origin of the end effector reference frame.

Unlike in the simulation, we tested the controller on a single task with different delay types.
\Cref{fig:hardware_rendering} depicts the 3D rendering of this task in rviz\footnote{\url{https://index.ros.org/p/rviz2/}}.
As a final result, our method successfully maintained safety during the teleoperation of the UR5 robot,
even when faced with various types of delays. \Cref{fig:ur5_details} shows the result when using recorded
network delays between Lund and Stockholm with an average RTT of \SI{11.61}{ms} and standard
deviation \SI{3.29}{ms}, indicating that the controller is capable of maintaining safe system behavior
also in real-time applications while under the influence of realistic delay.

\section{Conclusions}
\label{sec:Conclusions}
Our results indicate that
the proposed method effectively aligns with the objective of ensuring the safety of the controlled system by minimally modifying the input provided by a remote operator despite the network delays in a less conservative manner. 
In the case analyzed, safety refers to the ability of the system to avoid obstacles placed in the same environment where it has to operate.
The versatility of the method and the soundness of the implementation choices were validated through testing in both the Gazebo simulator and on a real UR5 industrial robot.
Future work may focus on the system's ability to maintain safety in scenarios involving packet loss, consider moving and differently shaped obstacles, and refine the accuracy of the system state prediction.

\section*{Acknowledgements}
The authors wish to thanks Matti Vahs, Jana Tumova, Matteo Saveriano, and Andrea Del Prete for their helpful feedback.

{
\bibliographystyle{IEEEtran}  
\bibliography{biblio}  
}

\end{document}